\documentclass[nofootinbib,aps,superscriptaddress,preprint,floatfix]{revtex4}

\usepackage{graphicx,epsf,subfig}
\usepackage{amsmath, amsthm, amssymb} 

\usepackage{bm}

\def\to{\rightarrow}
\newcommand{\nc}{\newcommand}

\nc{\beq}{\begin{equation}}
\nc{\eeq}{\end{equation}}
\nc{\barray}{\begin{eqnarray}}
\nc{\earray}{\end{eqnarray}}
\nc{\barrayn}{\begin{eqnarray*}}
\nc{\earrayn}{\end{eqnarray*}}
\nc{\bcenter}{\begin{center}}
\nc{\ecenter}{\end{center}}

\nc{\ket}[1]{| #1 \rangle}
\nc{\bra}[1]{\langle #1 |}
\nc{\mc}{\mathcal}
\nc{\er}[1]{(\ref{eq:#1})}
\nc{\onehalf}{\frac{1}{2}}
\nc{\partialbar}{\bar{\partial}}
\nc{\psit}{\widetilde{\psi}}
\nc{\Tr}{\mbox{Tr}}
\nc{\hc}{\mbox{H.c.}}
\nc{\ev}{\;\mathrm{eV}}
\nc{\mev}{\;\mathrm{MeV}}
\nc{\gev}{\;\mathrm{GeV}}
\nc{\tev}{\;\mathrm{TeV}}

\def\as{\alpha_S}

\def\chii0{\chi_i^0}
\def\chij0{\chi_j^0}

\newcommand{\gsim}{\lower.7ex\hbox{$\;\stackrel{\textstyle>}{\sim}\;$}}
\newcommand{\lsim}{\lower.7ex\hbox{$\;\stackrel{\textstyle<}{\sim}\;$}}
\nc{\ttbar}{t\bar t}
\nc{\la}{\langle}
\nc{\ra}{\rangle}
\nc{\ord}{{\cal{O}}}
\begin{document}
\begin{flushright}
IPPP/11/71\\
DCPT/11/142
\end{flushright}

\title{Structure of Fat Jets at the Tevatron and Beyond}

\author{Leandro G. Almeida \\
Institut de Physique Th\'eorique, CEA-Saclay, F-91191, Gif-sur-Yvette
cedex, France \\
Raz Alon\\
Department of Particle Physics and Astrophysics, Weizmann Institute of Science \\
Michael Spannowsky  \\ 
Institute for Theoretical Science, 
5203 University of Oregon, 
Eugene, OR 97403-5203, U.S.A. \\
and\\ 
Institute for Particle Physics Phenomenology, Department of Physics, \\ Durham University, DH1 3LE, United Kingdom
}

\begin{abstract}
Boosted resonances is a highly probable and enthusiastic scenario in any process probing the electroweak scale. Such objects when decaying into jets can easily
blend with the cornucopia of jets from hard relative light QCD states. We review jet observables and algorithms that can contribute to the 
identification of highly boosted heavy jets and the possible searches that can make use of such substructure information. We also 
review previous studies by CDF  on boosted jets and its measurements on specific jet shapes.
\end{abstract}

\maketitle

\section{Introduction}
Although the idea of looking inside a jet and study the radiation pattern of its constituents is not new \cite{Seymour:1993mx}, the large potential for searches of new electroweak scale particles has only been appreciated recently \cite{Abdesselam:2010pt}. At the LHC with its 14 TeV center of mass energy, particles with masses around the electroweak scale are frequently produced beyond threshold, i.e. boosted transverse to the beam direction. Either because they recoil against other energetic objects or because they arise from decays of even heavier particles, e.g. Z' or $KK$-gluons. If the resonance's transverse momentum is bigger than their mass, their decay products tend to be collimated in the lab frame. 

Jets, collinear sprays of hadrons and the hadrons' decay products, are the most frequent objects at the LHC. They do not only consist of final state radiation (FSR) but also of radiation from the Underlying Event (UE) and pileup. Therefore, it is important to be able to disentangle the rare events with boosted resonances from the large QCD background based on their different radiation patterns. Many different approaches have been developed to exploit these differences and will be 
the focus of this review. 
In the subsequent introduction we describe how jets are defined and the types of jets we are interested in, and how to properly understand the background in these kinematic regimes.
In Section \ref{sec:tooljet} we describe the different methods used to tag boosted jets through algorithms and observables that arise from the properties of the signal and background. In Section \ref{sec:meas} we describe a recent search at the Tevatron for boosted top jets and measurements of properties of highly boosted jets.

\subsection{Heavy Jets}
\label{sec:heavyjets}

It is not unambiguous what to call a jet in an event. To be able to compare experimental results with theoretical predictions, jets have to be defined in an infrared safe way. Therefore, IR safe sequential jet algorithms became increasingly popular over the last years. These algorithms sequentially merge (by combining their four-vectors) the pair of particles that are closest according to some distance measure $d_{ij}$ unless there is a distance $d_{iB}$ (so-called beam distance) which is smaller than all $d_{ij}$, in which case particle $i$ is called a jet and the algorithm proceeds with the remaining particles in the event. The most popular sequential jet algorithms are the $k_T$ \cite{Catani:1993hr, Ellis:1993tq} , the Cambridge/Aachen (CA) \cite{Dokshitzer:1997in, Wobisch:1998wt} and anti-$k_T$ algorithm \cite{Cacciari:2008gp}. Their measure $d_{ij}$ is defined by
\begin{equation}
  \label{eq:dij}
  d_{ij} = \min(p_{Ti}^{2n}, p_{Tj}^{2n}) \, \frac{\Delta R_{ij}^2}{R^2}\,,
  \qquad
  \begin{cases}
    k_T: \quad & $n=1$\,,\\
    \mathrm{C/A}:  \quad & $n=0$\,, \\
    \mathrm{anti-}k_T: \quad & $n=-1$,
  \end{cases}
\end{equation}

and $d_{iB}=p^{2n}_{T,i}$. Here, $R$ is the jet resolution parameter which specifies the size of the jet. These algorithms do not only ensure the comparability of Monte-Carlo (MC) calculations with experimental findings but by constructing the jet sequentially they also provide a recombination history of the jets' constituents. This history is attached to every jet in the event and allows to examine the jet at different resolution angles and intrinsic scales. Utilities for clustering and for studying the clustering history are available in Fastjet \cite{Cacciari:2005hq} and Spartyjet \cite{Ellis:2007ib}.

Different scenarios have been identified where jet substructure techniques are of value to improve phenomenological analysis. They can be cast into four different classes characterized by the resonance's $p_T$ and the overall business of the event.

\begin{description}
\item[large $p_T$, non-busy final state:] If the event is generated with a large invariant mass and the number of physical objects is small, e.g. a heavy TeV scale resonance decays into two electroweak scale resonances ($pp \rightarrow X_{\mathrm{TeV}} \rightarrow 2~Y_{\mathrm{EW}} \rightarrow \mathrm{jets} $) , the electroweak scale objects $Y_{\mathrm{EW}}$ tend to be highly boosted, their decay products are highly collimated and the radiation off the two $Y_{\mathrm{EW}}$ are well separated with respect to each other. Due to $p_{T,Y} \gg m_Y$ the fat jet's cone size does not need to be very large ($R \simeq 0.8$) to catch all the necessary FSR to reconstruct the boosted resonances. Because of the relatively small cone size and the absence of many sources of hard radiation in the event, the effect of UE, pileup, and ISR is less pronounced and one finds $m_j \simeq m_Y$. The importance of jet grooming methods to remove uncorrelated soft radiation is reduced but jet substructure methods are necessary to discriminate between a resonance jet and a QCD jet.
\item[large $p_T$, busy final state:] SUSY cascade decays are examples where, depending on the mass splitting of the SUSY particles in the decay chain, the boosted resonances can have large transverse momentum accompanied by many hard jets: $pp\rightarrow 2~X_{\mathrm{TeV}} \rightarrow Y_{\mathrm{EW}} +\mathrm{jets} \rightarrow \mathrm{jets} $. Considering UE, pileup, and ISR, the situation is similar to the non-busy final state, but the jet substructure methods might need to be adjusted to the many additional sources of hard uncorrelated radiation in the fat jet.
\item[medium $p_T$, non-busy final state:]
If two electroweak scale resonances are directly produced from proton collisions, $pp \rightarrow Y_{\mathrm{EW}} Y'_{\mathrm{EW}}$, they are usually produced around threshold, e.g. $pp\rightarrow HW$. Only a small fraction of the events yield $p_{T,Y} \gsim m_Y$.  Still, focusing on this fraction of events can be a superior way to disentangle the signal from the backgrounds. Phenomenological studies \cite{Butterworth:2008iy,ATL-PHYS-PUB-2009-088,Piacquadio:1243771,Brooijmans:2010tn,Plehn:2009rk,Soper:2010xk} have shown that, because of the kinematic features, event reconstruction efficiencies, b-tagging efficiencies \cite{ATL-PHYS-PUB-2009-088,Piacquadio:1243771} and the jet energy resolution can be improved and that combinatorial problems in the identification of the decay products of $Y_{\mathrm{EW}}$ are reduced. To make as many signal events accessible to jet substructure methods as possible the fat jet's cone size has to be much bigger than in the highly boosted scenarios. Therefore, UE, pileup, and ISR affect the fat jet's mass much stronger and one should consider using techniques to remove uncorrelated soft radiation, {\it e.g.} jet grooming methods~\cite{Abdesselam:2010pt}.
\item[medium $p_T$, busy final state:]
Events with many sources of hard radiation and less collimated decay products of the $Y_{\mathrm{EW}}$ are the most difficult scenarios, irrespective if jet substructure techniques are applied or not. Although many of the advantages outlined for the medium $p_T$ non-busy final states can be carried over, combinatorial problems to discriminate between decay products and hard radiation from other sources in the fat jet can occur much more frequently. Taggers for $Y_{\mathrm{EW}}$ have to take that into account and combine jet grooming techniques with criteria to discriminate $Y_{\mathrm{EW}}$ decay products from other hard radiation.
\end{description}

\subsection{Light QCD Jets}
\label{sec:lightjets}

The best way to understand how to distinguish heavy jet events initiated by tops, or other heavy states from 
the overwhelming light QCD jet background, i.\,e., jets initiated by hard light quarks and gluons, is to 
understand the behavior of jets in the kinematical regions of interest. In particular since we are dealing with 
high-$p_T$ jets, these should naturally have small R, or ``cone'' sizes, and thus allows to understand 
its analytical properties of its distributions versatilely. 

A particular useful feature of the hadronic jet cross-sections which we can exploit is its factorization properties. 
Factorization theorems provide us with a framework with which to compute such cross sections.
They allow to us to systematically separate long distance effects from short distance physics where perturbative 
QCD is applicable, i.\,e., when there is a large momentum transfer. Long distance effects are associated with 
the infrared regime of the theory, and thus are not perturbatively calculable. Together with a infrared-safe definition for the jets,
the cross-section for n-jet production factorizes into the following decomposition,
\begin{eqnarray}
	\frac{d \sigma_{ H_A H_B \to J_1 \dotsm J_n}(R_{def})}{ \prod_i^n d m_{J_i}^2 d p_{J_i,T} d eta_{J_i}} &=&  
	\sum_{abcd} \int  d x_a \, d x_b \, f^{H_A}_a(x_a,p_T) \, f^{H_B}_b(x_b,p_T)  \nonumber \\
	&& \hspace{-3cm} \times H^{IJ}_{ab\to cd} \left(x_a,x_b,\{ p^T_{J_i}, \eta_{J_i},   R_{def} \} \right) 
	S_{IJ} \left( \{m_{J_i}^2\,p^T_{J_i}, \eta_{J_i},R_{def} \}\right) \nonumber \\
	&& \hspace{-3cm} \times \prod_{i=1}^n J^{(c)}_{i} (m_{J_i}^2,p^T_{J_i}, \eta_{J_i},R_{def}).   \label{fact}
\end{eqnarray}
where $R_{def}$ are the parameters associated with the particular jet definition, such as the cone sizes or transverse momentum parameters. 

The functions $f^{H_{A,B}}_{a,b}$ are the parton distribution functions for partons $a,b$ in hadrons $H_{A,B}$.
$J_i$ are the jet functions, as the name implies are functions of only variables associatled to the particular Jet and its definition. 

The function $S_{IJ}$ represents the soft degrees freedom and describe wide angle radiation which correlates the final states, and thus also depends on the jet definition and jet observables, but depend only in the color structure of the states involved in the hard scattering.
The function $H^{IJ}$ describes the underlying hard perturbative process that produces the states that initiate the hadronic jets. It can be obtained from a partonic cross-section with the appropriate subtractions, the exact relations depends on the definition of the other factorized functions.

The jet function for quarks at fixed jet mass is defined by the following matrix element,
\begin{eqnarray}
J_{q_i} ( m_J^2, p_T,\eta, R^2)  &=& \nonumber \\ 
&& \hspace{-4cm}\frac{(2 \pi)^3}{2\sqrt{2} \, (p_{J}^0)^2}  \frac{ \xi_{\mu} }{N_c} \sum_{N_{J_i}} 
Tr \left\{ \gamma^\mu \la 0 | \Phi^{\dagger}_\xi ( 0, -\infty ) q(0) |N_{J_i}\ra \la N_{J_i}| \bar{q}(0) \Phi_\xi( 0, -\infty ) | 0\ra \right\} \nonumber \\
&& \hspace{-4cm} \times \delta \left( m_J^2 - \tilde{m}_J^2( N_{J_i}) \right))  \delta^{(2)} ( \hat{n} - \tilde{n}(N_{J_i})) \delta(  p^0_J - \omega(N_{J_c})
\end{eqnarray}
while gluon jets are defined as,
\begin{eqnarray}
 J_{g_i}( m_J^2, p_T,\eta, R^2)  &=& \nonumber \\ 
&& \hspace{-3cm} \frac{(2 \pi)^3}{2 (p_J^0)^3}   
\sum_{N_{J_i}}  \la 0 |  \xi_\sigma F^{\sigma \nu}  \Phi_{\xi} \left(0,\infty\right)  |  N_{J_i} \ra \la  N_{J_i} |\Phi_\xi^{\dagger}\left(0,\infty\right)   F_{\nu}{}^{\sigma} \xi_\sigma | 0 \ra
\nonumber \\
&& \hspace{-3cm}\times\delta \left( m_J^2 - \tilde{m}_J^2( N_{J_i}) \right)) \delta^{(2)} ( \hat{n} - \tilde{n}(N_{J_i})) \delta(  p^0_J - \omega(N_{J_c}).
\end{eqnarray}
They absorb collinear enhancements to the the outgoing particles in the underlying hard perturbative process. $\Phi$s are path ordered exponentials(Wilson lines) defined by 
$$\Phi_\xi (0,-\infty;0)= {\cal{P}} \left\{ e^{ -i g \int_{-\infty}^0 d \eta \, \xi \cdot A ( \eta\, \xi^\mu) }  \right\} $$
where $\xi$ is a direction with at least one component in the opposite direction of the jet,
the hadronic cross-section is independent of this direction. The jet function is normalized such that at lowest order its only contribution is,
\begin{equation}
 J^{(0)}_i ( m_{J_i}^2, p_T, R^2) =\delta( m_{J_i}^2 ).
\end{equation}
Contributions to the jet mass from the soft function only start contributing at $\ord(R^2)$. Thus in the case of small jets, $R<1$, most of the contribution are associated with
the jet function whose behavior starts $\ord(\log R^2)$. Note that the Soft and Jet functions are the only ones that depend on the particulars of the jet definition and thus contain all the substructure information. 

The rest of this discussion will deal only with high $p_T$ jets, since these are the jets that will have enough jet mass to be indiscernible from heavy jets initiated by tops or other electroweak processes, and thus are associated with small cone sizes. 
In this case the jet cross-sections for high $p_T$ and small cones simplify and analytical expression are rather manageable. At lowest non-trivial order, we have     

\begin{equation}
J_{i}=	\as (p_T)  \frac{  4\,  C_i}{\pi m_J} \log\left(\frac{R\,p_T}{m_J} \right)+ \ord(R^2),
\end{equation}
where $C_i$ are the color factors associated with the representation of the particle initiating the jet.
The contributions to the jet function from perturbative physics can be separated into 2 distinct
components.
\begin{figure}[h]
\centering
\includegraphics{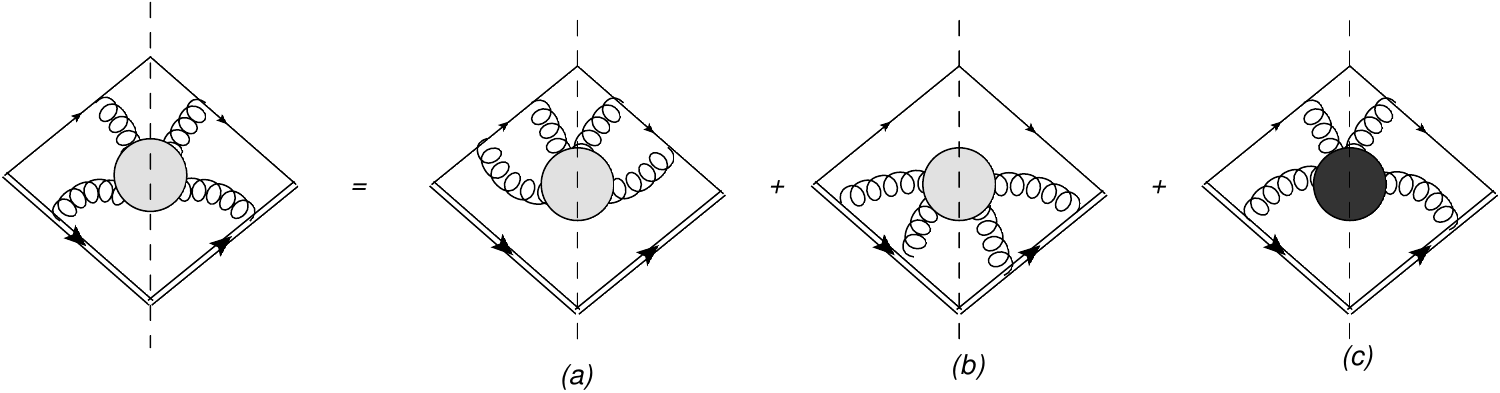}
\caption{Three seperate contributions to the Jet function in a general gauge}
\label{jetf}
\end{figure}
As shown in Figure \ref{jetf}, the first component constitutes of  contributions that only connect the quark lines. These are congruent to the splitting function contributions to the cross-section and differ only by a normalization constant. 

The second contribution connects only to the Wilson lines, this component can be made identically zero by making convenient choice of gauge. It is also possible, in the case of light like wilson line to choose their directions to give a similar cancelation. 
The third is simply the components that can not be separated into the first or second types, and thus are associated with incoherent interference contributions to the jet cross-section and affect the distribution of radiation around the jet. This component can usually be ignored for heavy particles that are produced on-shell and can be expanded in $\ord(\Gamma_p/m_p)$, where $\Gamma_p$ is the decay width of particle and $m_p$ its mass, e.g. the top quark or Higgs decaying into jets.

Thus for studying of the background jets, i.e. light QCD jets, and their substructure one may need to include at least components (a) and (c), while if one would like to focus on the substructure of high $p_T$ heavy particles, i.e. their radiation pattern, to $\ord(\Gamma_p/m_p)$ it's given completely by the splitting function contributions.

One interesting feature at leading order, is that the  jet mass distributions are independent of the properties of the other jets in the event. Therefore a measure of invariant mass distributions of each jet should provide a particular useful cut on the contamination of the light QCD jets in massive jet regions, such an observable is described in Section \ref{sec:meas}. Nonetheless such observables can be plagued by non-global logarithms, and further study of its high-order properties is warrant. Particularly since Monte Carlo studies \cite{Eshel:2011vs} have shown a slight correlation between the leading jet mass and sub-leading jet mass.

Beyond leading order one would need to include contributions from the soft function, whose definitions depend of the specific final states. In the case light QCD states it can be  simply described by the scattering of Wilson lines, however these would need to be properly subtracted to avoid double counting present from the current definitions of the jet functions. 
  
Jet filtering methods described below associated with removing soft radiation from jets with
a large cone size, algorithmic go through subjets and place cut precisely on their contributions from these jet functions. Since for heavy particles the only contributions are associated with the splitting functions, analysis of the ratios of energies and distances of substructure relative to the jet can be well understood at the leading log level. 




\subsection{Jet grooming methods}
\label{sec:cleanjet}

The purpose of the three methods called Filtering \cite{Butterworth:2008iy}, Pruning \cite{Ellis:2009su,Ellis:2009me} and Trimming \cite{Krohn:2009th} is to remove soft radiation from jets with a large cone size, e.g. $R=1.0$. Radiation from the UE or pileup tends to be soft, especially compared to the decay products of a boosted heavy resonance, but it has a big effect on the jet's mass $m_j$, i.e. the UE contribution to $\delta m^2_j$ increases with $R^4$.

Filtering and Trimming denote similar procedures. The fat jet's constituents are recombined with a smaller R which results in several subjets. 
Only a subset of these subjets are kept, 
based on a specific criterium which is often a $p_T$ threshold. 
The criterium to discriminate between hard and soft is different for Filtering and Trimming. While Filtering keeps a fixed number $n_f$ of the subjets, Trimming keeps only the subjets which fulfill

\begin{equation}
p_{T,j} > f \times \Lambda,
\end{equation}
where $f$ is an adjustable parameter and $\Lambda$ is an intrinsic scale of the fat jet, e.g. the fat-jet's transverse momentum. Note 
that although, 
$f$ is usually the same for all jets in the event but $\Lambda$ is not. Hence, the subjet $p_T$ threshold is different for every jet.

Pruning is a method which removes radiation from UE and pileup while clustering the fat jet. The sequential jet algorithm continues until no pair $(i,j)$ of constituents has 
a 
$R_{ij}>D_{\rm{cut}}$. $D_{\rm{cut}}$ represents an effective subjet cone size for this algorithm and is usually associated with an intrinsic fat jet scale, e.g. $D_{\rm{cut}}=m_j/p_{T,j}$. Now the protojet combination continues, but with an additional restriction: each pair $(i,j)$ of protojets that are due to be combined have to meet

\begin{equation}
z=\frac{\mathrm{min} (p_{T,i},p_{T,j}) } {| \vec{p}_{T,i} + \vec{p}_{T,j}|}>z_{\mathrm{cut}}.
\end{equation}
If $z<z_{\mathrm{cut}}$ the constituents $i$ and $j$ are not combined the one with smaller transverse momentum is discarded for the jet clustering. The algorithm continues until all constituents have either been combined or else eliminated.

For Trimming and Pruning the criteria to discriminate between FSR and UE/pileup adjust themselves based on quantities of the fat jet. To make Filtering work efficiently it is necessary to know the number of decay products of the resonance one seeks to reconstruct. Therefore Trimming and Pruning can be considered to be generic tagging algorithms for boosted resonances, particularly if other sources of hard QCD radiation are absent. 

It has been shown that all of these jet grooming techniques improve the jet's mass resolution when compared to the raw jet mass \cite{Abdesselam:2010pt}, e.g. the mass peak of a boosted top is narrower after applying grooming techniques. While the three methods perform very similarly in extracting the decay products of a hard resonance, they perform differently for QCD jets. This feature can be used to increase the significance of a resonance search \cite{Soper:2010xk}.
As long as the cuts are place on the physical properties of the subjets, such as jet momentum and its IR safe jet observables,
the Infrared Safety of these methods is congruent with the jet algorithm used to find the subjets. 
 
\section{Techniques to reconstruct boosted resonances}
\label{sec:tooljet}

A large set of resonance reconstruction techniques has been developed, 
most of which
focus on either reconstructing two pronged, e.g. Higgs boson, or three pronged decays, e.g. top quark, 
while some 
techniques are applicable to both decay patterns, e.g. N-subjettiness or energy-flow methods.

\subsection{Two pronged decays}
\label{sec:twoprong}

Depending on the $p_T$ of the resonance, the cone size of the fat jet has to be chosen in such a way that the resonance's decay products are contained in the jet to allow for a successful reconstruction. As a rule of thumb, this can be achieved by choosing $R \simeq 2 m/p_T$. Most subjet methods are based on the $k_T$ or CA algorithm. CA is a purely geometric algorithm which is not affected by the transverse momentum of the jets constituents, whereas the $k_T$ algorithm has a tendency to start with combining the softest constituents first and finishes in the last step with subjets which are hard and far apart, see Eq.(\ref{eq:dij}). Therefore, the $k_T$ algorithm has a natural tendency to revert the shower process when recombining the jet. While, the anti-$k_T$ algorithm provides a nice circularly shaped cone for the hardest jet in the event, it starts by combining hard jet constituents first, not following the naive shower picture. The anti-$k_T$ algorithm is therefore less suited for subjet methods than the $k_T$ or CA algorithms.

In an early study by Seymour \cite{Seymour:1993mx} it was shown that before detector effects were taken into account, clustering algorithms, a class that includes algorithms such as the $k_T$ algorithm, provided a much better reconstruction of the mass of a heavy boson decaying into the jet than conventional cone algorithms. A procedure now called ``YSplitter" was used in  \cite{Butterworth:2002tt, butterworth}, in which the authors used the $k_T$ algorithm to study the $d_{ij}$ distance in the final merging to reconstruct $W$ bosons. QCD backgrounds are likely to have small $d_{ij}$ values whereas $W$ jets tend to have values correlated with the mass of the $W$ boson.

Butterwoth, Davison, Rubin and Salam \cite{Butterworth:2008iy} proposed a method which combines filtering and a so-called procedure to reconstruct two pronged decaying resonances. This now goes by the name of ``BDRS method," and it makes use of the purely geometric substructure clustering of the CA algorithm. They suggested to undo several steps of the jet clustering until a substantial mass-drop emerges, e.g $\mathrm{max}(j_1,j_2)/m_{j_1,j_2} < \mu$, with $\mu$ substantially smaller than 1, and the transverse momentum of the subjets is required to be balanced, e.g. $\mathrm{min}(p_{T,j_1},p_{T,j_2})/\mathrm{max}(p_{T,j_1},p_{T,j_2})>y_{\mathrm{cut}}$. Wide-angle UE and pileup radiation is actively removed in this process but the two resulting subjets after meeting the mass-drop condition might still contain uncorrelated soft radiation. This radiation is removed in a following step by applying Filtering on the constituents of the two subjets. The constituents are reclustered with CA, using $R=\mathrm{min}(0.3,\Delta R_{j_1,j_2}/2)$ and only the three hardest of these subjets are taken. The third subjet is taken not to miss the soft wide-angle radiation off the decay product's color dipole. This method proofed to be successful in reconstructing the Higgs, Z and W bosons masses without introducing artificial scales in the QCD jets' mass distribution.

Variations of the BDRS method have been exploited in studies with the purpose to reconstruct the Higgs boson in a busy final state, e.g. $t \bar{t} H$ \cite{Plehn:2009rk} or SUSY cascade decays \cite{Kribs:2009yh,Kribs:2010hp}. In busy final states the two resulting subjets after meeting the mass drop condition might not be directly correlated to the decay products of the resonance. Therefore, it can be advantageous not to stop after the mass-drop condition is met for the first time, but to keep undoing the jet clustering 
while monitoring 
the structure of each individual mass drop. After going through the whole tree the two relevant subjets can be selected based on a well suited subjet criterium. 

\subsection{Three pronged decays}
\label{sec:threeprong}

The most prominent example of a three pronged decay used for subjet studies is the decay of a top quark:
\begin{equation}\label{eq:top_decay}
t \rightarrow W^+ b \rightarrow q'\bar{q}b.
\end{equation}
Initially studying boosted tops was motivated by the potential decay of a very heavy resonance ($m_x \simeq \mathcal{O}(1)~\mathrm{TeV}$) into tops \cite{Agashe:2006hk,Lillie:2007yh,Baur:2007ck,Baur:2008uv}, leading to highly collimated three-pronged decays ($p_{T,\mathrm{top}}>500~\mathrm{GeV}$). The two tops recoil against each other and the reconstruction of the heavy resonance, e.g. a $Z'$ or $KK$-Gluon, is jeopardized because the top decay products can overlap and have to overcome large QCD dijet or $W$+jets backgrounds. Jet substructure methods are unavoidable under these circumstances. The fact that the decay products of the two tops are well separated and the knowledge of the top and W mass can be used to develop so-called top taggers with good reconstruction and background rejection efficiency. Often top taggers are inspired by the ideas outlined in Sec.(\ref{sec:twoprong}). The performance of several of the taggers described in this section were compared in \cite{Abdesselam:2010pt}.

Thaler and Wang \cite{Thaler:2008ju} utilize a similar approach. A jet is reclustered with the $k_T$ algorithm, until two or three subjets are formed. Internal kinematic variables in addition to $k_T$ scales are combined to disentangle the top jet from the background. For example, in the three-subjet analysis, a W boson candidate is identified by forming the minimum pairwise mass between subjets and a minimum cut is placed on its mass. Relative energy sharings between the subjets 
were also studied.

A top tagger early applied by ATLAS \cite{Brooijmans:1077731,ATL-PHYS-PUB-2009-081} is an extension of the ``YSplitter". Using the $k_T$ algorithm, the $d_{ij}$ values of the next-to-last and next-to-next-to-last clusterings are combined with the jet's mass in a multidimensional space, which yields enough information to apply cuts or a likelihood ratio to discriminate between top jets and QCD jets. The tagger performs particularly well for large tagging efficiencies \cite{Abdesselam:2010pt}.

The ``Hopkins" top tagger is based on the mass-drop procedure of the BDRS study \cite{Kaplan:2008ie,hopkinstaggerURL}. The fat jet is clustered using the CA algorithm. This jet is 
then 
decomposed by reversing the cluster history until a branching is 
found, such that 
the $p_T$ of the subjets is balanced and the distance between the two subjets is not too small. This procedure is repeated on the two subjets. A jet is considered a top-jet candidate if after the first mass-drop at least one more mass-drop was found. On the resulting 3 or 4 subjets several kinematic cuts are aplied: The invariant mass of the all subjets should lie near $m_t$, two of the subjets are required to reconstruct new $m_W$ and their helicity angle should not be too small.

The ``Hopkins" top tagger has been modified by CMS \cite{cmsnote3}, 
where the  
W boson mass window cut and the helicity angle cut have been replaced by a single cut on the minimum pairwise subjet mass.

All of these taggers are focusing on reconstructing top quarks with large transverse momentum, therefore none of these taggers applies a jet grooming procedure like Filtering, Pruning or Trimming, which makes them more sensitive to the fat jet's cone size. 

The Heidelberg-Eugene-Paris (HEP) top tagger \cite{Plehn:2009rk, Plehn:2010st} was designed to work in the medium $p_T$ region for busy and non-busy final states, i.e. with very large 
jet radius $R=1.5$. Inspired by the BDRS method, the tagger uses a combination of mass drops and Filtering. While branching through the whole 
jet's recombination history down to a subjet-mass scale of 30 GeV. In every merging very soft subjets are discarded. To select the decay products of the top quark a minimization procedure of the remaining hard substructure is applied. Every set of constituents of the three subjets is reclustered with CA, $R = \mathrm{min}(0.3, (\Delta R_{ij}/2))$ and $n_f=5$. The set of three initial subjets that gives a Filtered mass closest to the top mass is retained as top candidate. This top candidate is reclustered again to yield exactly 3 subjets, which are the anticipated decay products of the top quark. The invariant-mass combinations of the subjets are placed in the two-dimensional subspace of $(m_{23}/m_{123},\mathrm{arctan}(m_{13}/m_{12}))$. For tops, one of the combinations is required to satisfy $m_{ij}/m_{123} \simeq m_W/m_t$. To further increase the efficiency of the tagger, dipolarity \cite{Hook:2011cq} was proposed to single out the decay products of the W.

Three pronged decays can also occur in searches of new physics. In \cite{Butterworth:2009qa}, the authors explore two tagging methods in R-parity-violating neutralino decays. One method is similar to the YSplitter, cutting on dimensionless ratios as $d_{12}/m^2$. The other method is CA based and searches the entire cluster history of the fat jet. For every merging which is not too asymmetric in $p_T$ the Jade-distance $p_{T1}p_{T2} \Delta R_{12}^2$ is recorded. The clustering with the largest Jade-distance defines the neutralino candidate and to ensure 3-body kinematics, a cut is placed on the ratio of the masses of the subjets with second-largest and largest Jade distances.

\subsection{Jet-shape and energy-flow methods}
\label{sec:jetshape}

Infrared safety of observables guarantees that we can make meaningful comparisons between theoretical computations and experimental measurements. Such observables should not probe regions directly associated with infrared singularities of the theory, since within these regions the perturbative cross-sections are ill-defined. These regions are associated with collinear and soft degrees of freedom, therefore infrared-safe observables should be insensitive to collinear splittings of states or additional soft momenta. 
These observables should be insensitive to non-perturbative physics, since these are the same regions are associated with long distance effects that are not a priori computable in the framework perturbative QCD.

Such observables can be built from information of the states within the jet. We call such IR-safe observables, 
jet shapes, since they should be sensitive to the distribution of energy and momenta within the jet.
There have been a variety of jet shapes introduced, some are simply applications of previous event-wide observables,
i.e. event-shapes, applied to the subset of event states that clustered into a jet. Therefore they must be defined 
for any number of states that may define the jet.

One such observable used in the identification of massive jets is the so-called Planar flow proposed 
in \cite{Thaler:2008ju, Almeida:2008yp,Almeida:2008tp}. For example, in \cite{Almeida:2008yp,Almeida:2008tp},
one begins by defining a tensor $I_{\omega}$ that depends on the energy of the states defining the jet and their 
momentum transverse to the direction of the jet momentum, $p_{i,k}$. 

\begin{equation}
\label{eq:planar_tensor}
I_\omega^{kl}=\frac{1}{m_J} \sum_i \omega_i \frac{p_{i,k}}{\omega_i} \frac{p_{i,l}}{\omega_i},
\end{equation}

One can then construct an IR-safe observable from the eigenvalues of this tensor, such as the  
one defined in \cite{Almeida:2008yp,Almeida:2008tp},
\begin{equation}
	P_f = 4 \frac{\det I_{\omega}}{(Tr I_{\omega})^2}.
\end{equation}
The interesting the feature of this observable is that it's zero for any linear distributions in the plane transverse 
to the jet axis. Therefore, if the jet decay is dominated by single emission from the particle initiating the jet, 
then its Planar flow distribution should peak around zero. Infrared safety of this observable makes it less sensitive 
to hadronization effects, and therefore perturbative features such as low planar flow are expressed at the hadron level.

It was first used in distinguishing high-$p_T$ top decays as in Eq. \ref{eq:top_decay}, where the background, high mass light-QCD 
jets, is dominated by single gluon emission. Therefore it greatly differentiates light-QCD jets from top jets, whose decay 
chain leads to a flat Planar flow distribution.

Another use of Planar flow in \cite{Chen:2010wk,Falkowski:2010hi,Almeida:2010pa}
where it was used in a contrasting manner, not to find background events with low planar flow as above, 
but to find signal events with lower planar flow than the background. 
If we are interested in looking for color neutral jets, then most of the radiation is exchanged 
among the subsequent QCD final states in the decay chain, since at the level of leading color the 
radiation is color blind to rest of the event. Angular ordering of soft gluon emission ensures 
that more radiation off a color dipole can be found between the color connected partners \cite{Marchesini:1983bm}.
This leads to a linear distribution of high final states in-between two hard events within the jet. 
Therefore, Planar flow distribution for electro-weak boson jets, or other exotic color neutral jets, 
will peak at lower planar flows then QCD jets. The method should be particular useful for high$-p_T$
regions where light-QCD jets are contaminated by extra radiation, i.e. where single gluon emission doesn't dominate.

The use of the energy distribution between two jets or subjets to distinguish the color of the final states in an event was discussed in \cite{Sung:2009iq,Gallicchio:2010sw}. In \cite{Gallicchio:2010sw} an observable called ``pull'' was developed, it makes use of the gradient of the transverse energy distribution for each of jets in an event. The angle between the gradients provides a measure of the color pertaining to the original state that decayed into the two jets (or subjets). This observable has shown great promise in eluding the color structure of not only in jets themselves but also event-wide.

Another method to distinguished heavy jets using their energy distributions was introduced in \cite{Almeida:2010pa}. The authors began by trying to obtain the utmost distinction of two jets by the difference in their energy distributions, by introducing a method to match their IR-safe energy distributions between fixed order energy flow calculations and the measured energy distributions. The measured energy distributions can be match to a set of fixed templates, which describe the kinematical information from signal or background. It proved particularly useful when combining with cuts in jet shapes, such as Planar flow, or in the kinematics of the matching set of templates. The method itself is versatile enough to work for a range of processes, it is particular useful for events where the energy distribution is all that is available.

%

%
%
%

Recently, N-Jettiness has been introduced, an event shape observable which is designed specifically to facilitate the theoretical description of exclusive N-jet final states. Based on \cite{Stewart:2010tn}, in \cite{Kim:2010uj} and \cite{Thaler:2010tr,Thaler:2011gf} the approach was adapted to construct a jet shape observable N-subjettiness $\tau_N$. 
$\tau_N$ is calculated by minimizing
\begin{equation}
\tau_N=\frac{1}{d_0} \sum_k  p_{T,k} ~\mathrm{min}  \left \{ \Delta R_{1,k} \cdots \Delta R_{N,k} \right \},
\end{equation}
with
\begin{equation}
d_0 = \sum_k p_{T,k}R.
\end{equation}
Here the index $k$ runs over all constituents of the fat jet with cone size $R$. By minimizing $\tau_N$, N axes of candidate subjets are constructed. $\tau_N$ vanishes if the number of subjet candidates coincides with the number of jet constituents. Therefore, $\tau_N$ can be used to measure the number of hard isolated energy deposits inside a fat jet which can be helpful to disentangle resonance jets from QCD jets. To ameliorate the effect of hard uncorrelated radiation in the fat jet on $\tau_N$, it can be advantageous to consider $\tau_N/\tau_K$, with $K \neq N$. 

Looking for mass-drops while branching backwards through the jet recombination tree is a recurring motif in many resonance reconstruction techniques. An approach avoiding trees was proposed in \cite{Jankowiak:2011qa}. The angular correlation function
\begin{equation}
\mathcal{G}(R)=\frac{\sum_{i \neq j} p_{Ti} p_{Tj} \Delta R^2_{ij} \Theta(R-\Delta R_{ij})}{\sum_{i \neq j} p_{Ti} p_{Tj} \Delta R^2_{ij} },
\end{equation}
where the sum runs over all pairs of constituents of the jet and $\Theta(x)$ is the Heaviside step function, provides a profile of the jet radiation. The variation of $\mathcal{G}$ with $R$, $d \log \mathcal{G}(R)/d \log R \geq 0$, yields a topographic profile of the jet. In \cite{Jankowiak:2011qa} the hight and number of peaks is used tag top-quark jets.

\subsection{Leptonic top quark reconstruction}
\label{sec:leptop}

On the one hand, the lepton, the b-jet and the sizable amount of missing transverse energy provide strong handles to suppress backgrounds if the signal contains a leptonic top. On the other hand, the invisible neutrino momentum and the difficulty of measuring the neutrino's transverse energy precisely, hampers the top quark's reconstruction. 
If the top quark is boosted  to very large $p_T$, standard lepton isolation criteria reduce the signal efficiency and b-tagging can be severely degraded.
Using a ``mini-isolation" cut \cite{Rehermann:2010vq} at the tracker level for muons is a tempting option to reject a large fraction of background events while retaining a good signal reconstruction efficiency. Alternatives to lepton isolation cuts have been proposed in \cite{Thaler:2008ju} making use of kinematic correlations of the top decay products.
If the lepton is isolated, knowing the neutrino momentum ensures a good reconstruction of the top's momentum. In many searches the neutrino momentum is reconstructed by assuming that all missing transverse energy origins in the neutrino momentum, in which case the neutrino can be reconstructed by requiring $m_{\nu l}=m_W$. In \cite{Plehn:2011tf} a novel way of reconstructing the neutrino momentum was introduced, using only the lepton and b-jet momenta. By not relying on the missing transverse energy vector the method is less sensitive to systematic uncertainties and can be applied to searches where more sources of MET are present, e.g. $pp\rightarrow \tilde{t} \bar{\tilde{t}} \rightarrow t \bar{t} +\mathrm{MET}$. 

\subsection{Maximum information approaches}

The outlined jet substructure methods apply different approaches and ideas to disentangle the sought hard substructure associated with the hard interaction of the event from QCD backgrounds. Hence, the different methods might extract complementary information with respect to each other and it can be beneficial to combine them to obtain a better signal purity or background rejection \cite{Soper:2010xk}. 

In \cite{Gallicchio:2010dq} the authors combined many jet and event shape observables to improve the significance for the $HV$ search with $H\rightarrow b\bar{b}$ at the Tevatron. These observables were processed and analyzed using Boosted Decision Trees to increase the statistical significance over cut based studies. Systematic uncertainties are difficult to asses, but a significance increase of $10-20 \%$ was found compared to the variables used by the CDF and D0 collaborations.
A similar approach was used for the analysis of boosted W in \cite{Cui:2010km}, where grooming techniques and jet substructure observables like R-cores were combined to improve on the BDRS method. After processing them through a Boosted Decision Tree, a significance improvement for highly boosted W over the BDRS method was found.

The method ``Shower deconstruction" \cite{Soper:2011cr} aims to be a maximum information approach by acting on the smallest practical pieces of radiation in an event, e.g. topo-clusters, calorimeter towers or just small jets. Shower deconstruction then assigns to each event a number obtained from an analytic function that is an estimate of the ratio of the probability for a signal process to produce that event to the probability for a background process to produce that event. The analytic function to calculate these probabilities is based on first principle QCD and mimics what full event generators do. The method was applied to the $HZ$ final state and was found to improve on the BDRS study in this channel.


\section{Searches and Applications}
\label{sec:searches}

The existence of boosted resonances is a kinematic necessity if electroweak scale physics is probed at a multi-TeV collider. The reconstruction of boosted electroweak scale objects can be useful in a plethora of searches for new physics. Thus, the jet substructure tools outlined in Sec.\ref{sec:tooljet} have been used and developed for this purpose. Most new physics scenarios propose heavy particles which decay subsequently in electroweak scale particles. Therefore, searches which make use of jet substructure can provide a superior way to discover new physics by reconstructing boosted particles like electroweak gauge bosons, the Higgs boson or the top quark.

\subsection{Boosted electroweak gauge bosons}

The first applications of jet substructure techniques were focusing on the reconstruction of boosted hadronically decaying gauge bosons \cite{Butterworth:2002tt}. Studying longitudinal vector boson scattering can provide insights into the nature of electroweak symmetry breaking, especially if no Higgs boson is found. For this purpose, the vector boson fusion process, where two electroweak gauge bosons are produced in association with two tagging jets, is particularly suited. Standard searches rely on leptonic boson decays. Allowing them to decay hadronically can increase the sensitivity of the study. In \cite{Han:2009em} this was further treated using the polarization of the vector bosons.

SUSY cascade decays are a source of boosted electroweak gauge bosons. In \cite{Butterworth:2009qa} their reconstruction was used to obtain information about masses and branching ratios of SUSY particles.

A heavy Higgs boson ($m_h>300~\mathrm{GeV}$) decays most of the time into gauge bosons. Its main discovery channel, the so-called ``gold plated mode", relies on two boosted leptonically decaying Z bosons. This channel has only small backgrounds and provides a very clean signature with a good $S/B$. However, using the BDRS method improved by a combination of Trimming and Pruning \cite{Soper:2010xk}, even the semihadronic channel was found to be of comparable significance as the purely leptonic channel, despite its much bigger backgrounds \cite{Hackstein:2010wk}. 
The semihadronic channel also provides sensitivity to the CP property of a heavy scalar resonance \cite{Englert:2010ud}. Generically, the decay of very heavy resonances, e.g. $Z'$, can result in highly boosted Higgs boson or electroweak gauge boson which have a sizable branching ratio into bottom quarks or taus. This scenario is discussed in \cite{Katz:2010mr, Katz:2010iq}.

\subsection{Boosted Higgs bosons}

Hadronic decays of the Higgs boson, e.g. $h\rightarrow b\bar{b}$, have played an important role in searches at the Tevatron. At the LHC, due to an increased Higgs-production cross section and a simpler reconstruction, most of the searches rely on decays to gauge bosons or leptons. The Higgs boson decays into a collimated $b\bar{b}$ pair with a large branching fraction. The analysis \cite{Butterworth:2008iy, ATL-PHYS-PUB-2009-088} makes use of the BDRS method outlined in Sec.\ref{sec:twoprong}. It demonstrates that the VH production channel can be a viable candidate for the discovery of a light Higgs boson at the LHC. The ATLAS collaboration reproduced this result in a full-detector simulation, finding $S/ \sqrt{B} \simeq 3.7$ after $\mathcal{L}=30~\mathrm{fb}^{-1}$ considering only statistical errors. 

In early ATLAS and CMS reports \cite{:1999fr,Drollinger:2001ym,Abdullin:692492} the $t \bar{t}H$ production channel with subsequent Higgs decay to bottom quarks was one of the major discovery channels for a light Higgs boson. Further studies revealed a very poor signal-to-background ratio of 1/9 \cite{Cammin:685523}, making the channel very sensitive to systematic uncertainties which might prevent it from reaching a $5 \sigma$ significance for any luminosity. However, at high transverse momentum, after reconstructing the boosted, hadronically decaying top quark using the HEPTopTagger as well as the Higgs boson with a modified version of the BDRS method, and requiring 3 b-tags, the signal-to-background ratio can be improved to $\sim1/2$, while keeping the statistical significance at a similar value to that in Ref. \cite{Cammin:685523}.

If the scalar sector is more complex than in the SM predicted, the Higgs boson can decay into light pseudo-scalars $a$ and possibly evade constraints from the LEP experiments. If $m_a \ll m_h$ the light scalars are boosted and the decays to gluons \cite{Chen:2010wk, Falkowski:2010hi}, light quarks or taus \cite{Englert:2011iz} can dominate. All of these studies exploit that the radiation pattern of a color singlet state differs from QCD backgrounds. They conclude that the Higgs boson can be reconstructed with as little as $\mathcal{L}=20-100~\mathrm{fb}^{-1}$. The lightest CP-even Higgs boson of the MSSM might even be discovered earlier than in any other channel, if searched for in Squark or Gluino cascade decays \cite{Kribs:2009yh,Kribs:2010hp}.

\subsection{Boosted top quarks}

The need to be able to reconstruct boosted tops from a heavy resonance \cite{Agashe:2006hk,Lillie:2007yh,Baur:2007ck,Baur:2008uv} or top partner decay \cite{Plehn:2010st,Plehn:2011tf,Kribs:2010ii} has motivated and promoted the field of jet substructure like anything else. Top quarks from heavy resonances, e.g. $Z' \rightarrow t \bar{t}$, are naturally boosted with $p_{T,t} \gtrsim 500~\mathrm{GeV}$. While top quarks produced around threshold yield well separated decay products and can be efficiently discriminated from QCD jets, at a first glance, a highly boosted top looks like one single massive jets. Jet substructure methods are essential to discriminate between QCD and top jet. 
Because of the technical difficulties in reconstructing boosted tops a lot of work has been devoted to developing tools, whereas the number of papers applying these tools to new physics searches is still relatively small - yet growing.

The �Hopkins� top tagger  has been used in phenomenological studies reconstructing KK  vector boson resonances \cite{Bhattacherjee:2010za} and full detector simulation studies have been performed by CMS with a slightly modified implementation \cite{cmsnote3}. For the reconstruction of a heavy $Z'$ \cite{CMS-PAS-EXO-09-002} between 1-4 TeV they find a good background rejection and signal efficiency. Using the ``Hopkins" top tagger can increase the sensitivity for $Z'$ with $m_{Z'}>2~\mathrm{TeV}$ compared to standard reconstruction techniques.

Because scalar top partners can ameliorate the top quark's impact to the hierarchy problem of the Higgs, they are among the most anticipated particles to be found at the LHC \cite{Meade:2006dw}. The HEPTopTagger was applied \cite{Plehn:2010st} to reconstruct the light top squark of the MSSM in a final-state with only jets and missing transverse energy, $pp\rightarrow \tilde{t} \bar{\tilde{t}} \rightarrow t \bar{t}  \chi^0_1 \bar{\chi}^0_1 $. While a reconstruction with standard techniques yields $S/B \simeq 1/7$, a subjet analysis in combination with $m_{T2}$ \cite{Lester:1999tx, Barr:2002ex} can result in $S/B\simeq 0.88$ and $S/\sqrt{B} \simeq 6$ after $10~\mathrm{fb}^{-1}$. If one of the tops decays leptonically a leptonic top tagger can be used to separate the neutrino's MET contribution from the neutralinos' MET, which allows an effective use of $m_{T2}$. The significance in the channel with semileptonic top decays is similar to the all-hadronic channel \cite{Plehn:2011tf}. 

The top can be produced from a top partner decay in association with a Higgs boson. Even if the top and the Higgs are not highly boosted they can be reconstructed using the BDRS approach and the HEPTopTagger, with only $10~\mathrm{fb}^{-1}$ at 14 TeV \cite{Kribs:2010ii}.

%


\section{Measurements}
\label{sec:meas}
In this section we discuss the experimental measurements of boosted jets at CDF.
Experimental studies at the Tevatron were limited to jets with $p_T < 400~\rm{GeV}$ \cite{Abazov:2001yp, Acosta:2005ix}. Results on jets with higher $p_T$ produced at the LHC have been published recently~\cite{Aad:2011kq}.

The CDF study presented here~\cite{Aaltonen:2011pg, CDF:subjet2010a} was the first to look at substructure of massive jets with $p_T > 400~\rm{Gev}$. Both Midpoint and anti-$k_T$ jet algorithms were used, with a typical jet size of $R = 0.4, 0.7,$ and $1.0$. A pileup correction technique was used to remove contribution of non-coherent energy deposits, critical in substructure studies. Three observables were measured. The jet mass was compared to the theoretical next-to-leading order (NLO) prediction of the jet function~\cite{Almeida:2008tp}. Angularity and planar flow~\cite{Almeida:2008yp} served as further support to the hypothesis of the two-prongness of QCD jets.  Finally, a boosted top search~\cite{CDF:subjet2010b} based mostly on the jet mass will be reviewed.

\subsection{Pileup Correction}
Pileup is the result of additional collisions in the same bunch crossing. The number of reconstructed vertices, $N_{vtx}$, is a good estimate for the number of multiple interactions. Its average in the studied data was $\sim 3$. Pileup creates additional energy deposits in the detector which are crucial for substructure studies. This problem becomes more severe at the large luminosities expected at the LHC. Moreover, the pileup in the MC samples used is not necessarily modeled exactly as the data. Therefore a data driven technique was used to measure the shift in various observables due to additional incoherent energy in the jet and correct for it~\cite{Aaltonen:2011pg, Alon:2011xb,Sapeta:2010uk}.
A complementary cone, one with the same size parameter $R$, is set at $90^o$ in azimuth away from the leading jet. The energy deposits inside this cone are rotated back and added to the jet. Then the shift in the interesting observables can be measured as a function of the value of these observables. Due to the large event-by-event fluctuation in this energy, the mean shift was used for correction instead of an event-by-event value. Furthermore, since the underlying event contributes some coherent energy to the final state, its contribution was separated from that of the multiple interactions by the following method: the shift from single-vertex events (assumed to contain only underlying event contributions) was subtracted from the shift in multi-vertex events (containing both underlying event as well as multiple interaction contributions).

This technique was implemented on the jet mass, reducing the systematic shift in the jet mass distribution due to pileup. Finally, an analytic calculation for deriving an approximation to these shifts in the observables is presented in~\cite{Alon:2011xb}, and the relation with the concept of jet area is discussed.

\subsection{Jet Mass}
The jet function~\cite{Almeida:2008tp} predicts both the shape of the jet mass distribution, given its functional form, as well as the absolute normalization. In the reviewed study, an NLO approximation was used, valid for jet masses above the low mass peak and below the value of $R\cdot p_T$ where $R$ is the typical size used to associate particles to a jet. Higher-order corrections were estimated to be $\sim 30$\%.

\begin{figure}[h]
\centering
\includegraphics[width=8cm]{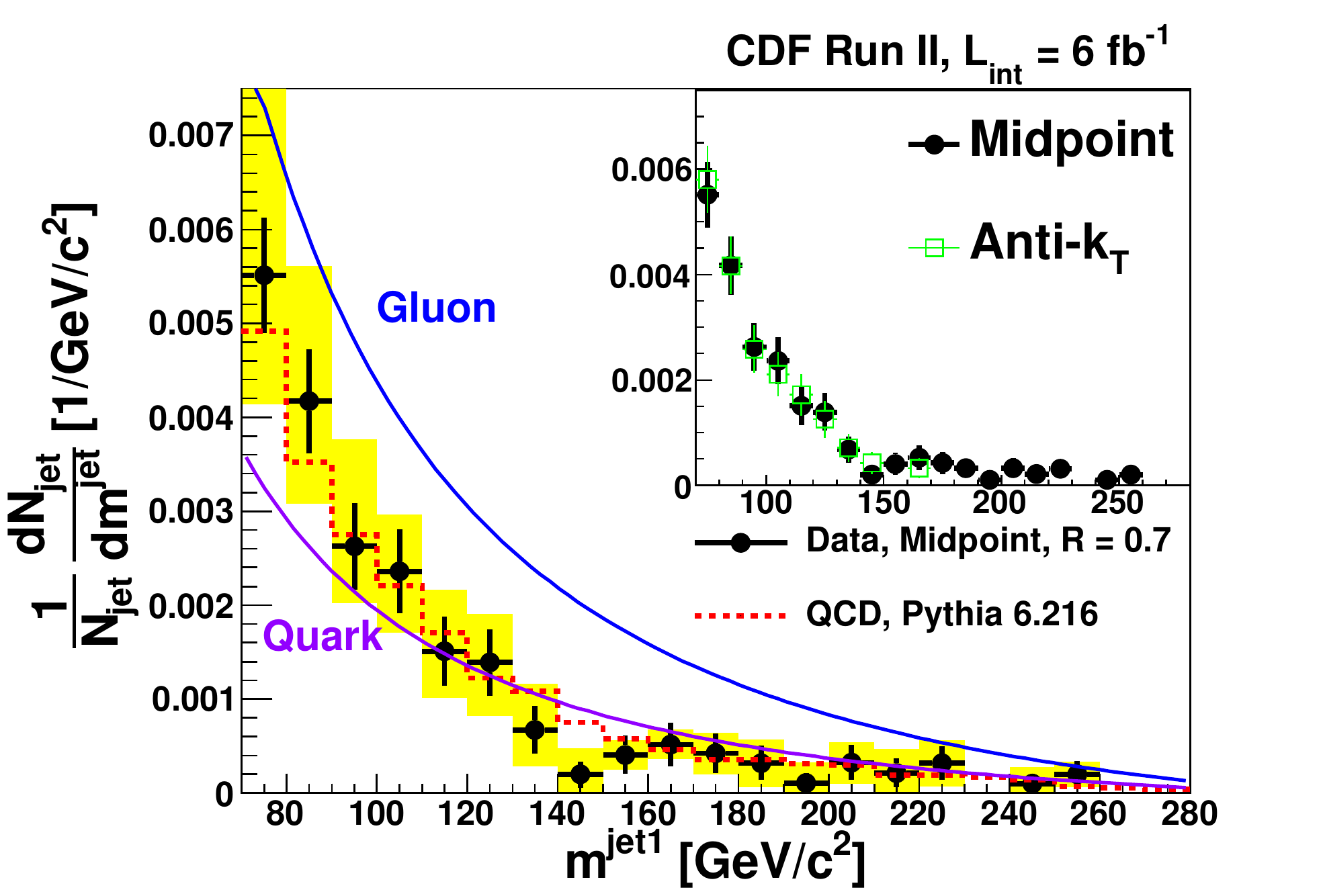}
\caption{The normalized jet mass distribution for Midpoint jets with $p_T > 400~\rm{Gev}$ and $|\eta| \in (0.1, 0.7)$. The uncertainties shown are statistical (black lines) and systematic (yellow bars). The theory predictions for the jet function for quarks and gluons are shown as solid curves and have an estimated uncertainty of $\sim 30$\%. Also shown is the PYTHIA MC prediction (red dashed line). The inset compares Midpoint (full black circles) and anti-$k_T$ (open green squares) jets~\cite{Aaltonen:2011pg}}
\label{massC7}
\end{figure}

Fig.~\ref{massC7} shows a comparison of the jet mass distribution for a cone size R = 0.7 with the analytic predictions for the jet function. To conform to the validity range of the jet function, this comparison is made for jet masses above $70~\rm{Gev}^2$ and up to $280~\rm{Gev}^2$. The analytical prediction for quark jets describes approximately the shape of the distribution and fraction of jets but
tends to over-estimate the rate for jet masses from $130$ to $200~\rm{Gev}^2$. pQCD indeed predicts that $\sim 80$\% of these jets arise from quarks~\cite{Ellis:1991qj}. \\
The data and the PYTHIA distributions are in reasonable agreement. Furthermore, as shown in the inset plot, a good agreement is observed between the Midpoint and the anti-$k_T$ algorithms. Since the jet mass is and NLO effect (a gluon emission is needed) this is very interesting considering the fact that Midpoint is shown to be IR$_{3+1}$~\cite{Salam:2009jx} whereas anti-$k_T$ is IR-safe.

\subsection{Angularity}
Angularity was shown to qualitatively distinguish between QCD jets and other two-body decays~\cite{Almeida:2008yp}. The reviewed study showed that the shape of the angularity distribution agrees with the two-prong description of high $p_T$ massive QCD jets, an assumption that the jet function is also based upon. This agreement is manifested in the two kinematical limits: $\tau_{-2}^{min} \sim (m_j/2p_{T,j})^3$ is obtained from decay configurations in which both daughter particles are emitted at the same angle with respect to the direction of the mother particle and have the same energy. $\tau_{-2}^{max} \sim 2^{-3}R^2m_j/p_{T,j}$ is obtained when one of the decay daughters is hard and almost collinear with the mother particle, whereas the second decay daughter is soft and emitted at a large angle, limited by the size parameter $R$.

\begin{figure}[h]
\centering
\includegraphics[width=8cm]{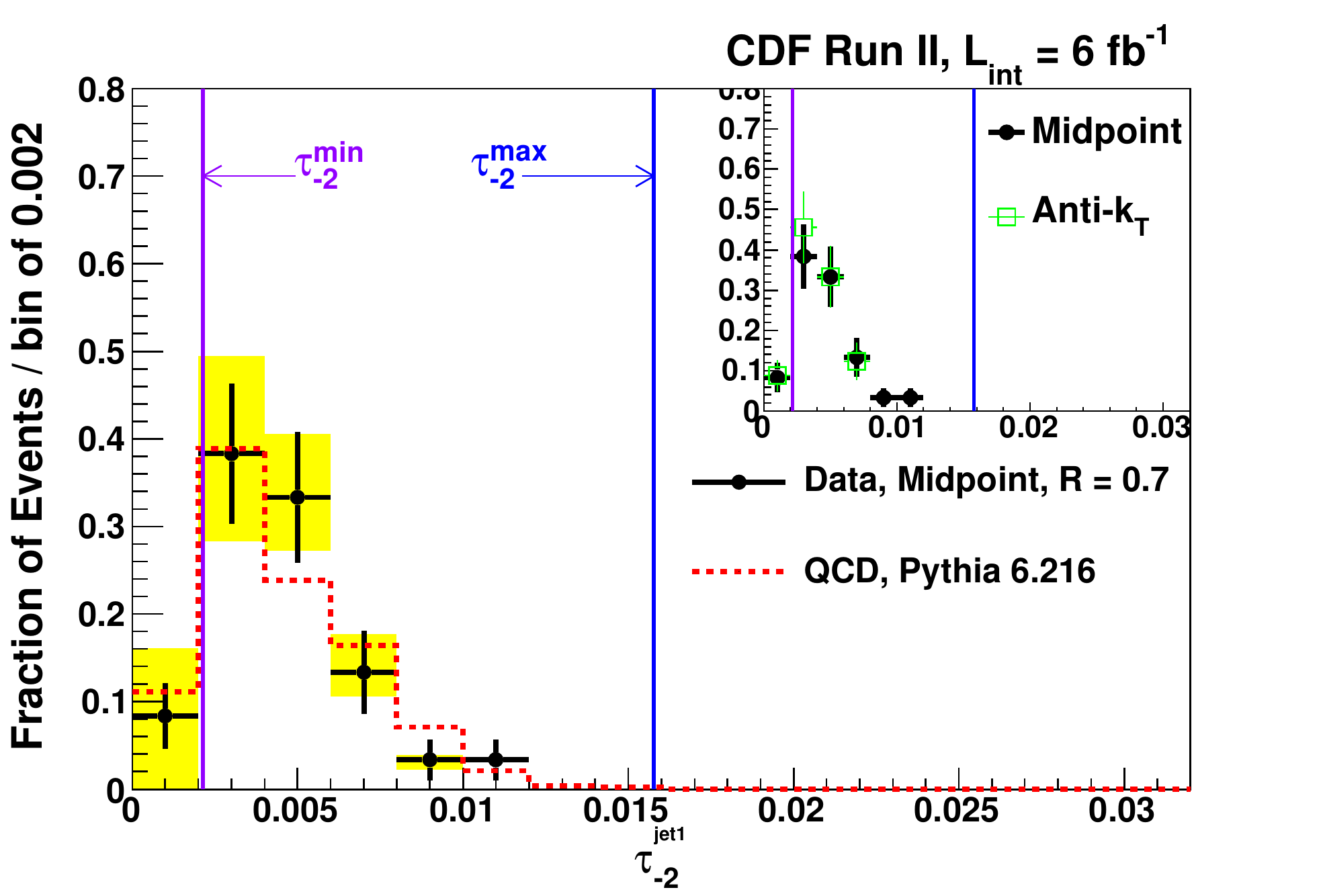}
\caption{The angularity distribution for Midpoint jets with $p_T > 400~\rm{Gev}$, $|\eta| \in (0.1, 0.7)$, and $m^{jet1} \in (90, 120)~\rm{Gev}^2$. Also shown are the PYTHIA calculation (red dashed line) and the pQCD kinematic endpoints. The inset compares the distributions for Midpoint (full black circles) and anti-$k_T$ (open
green squares) jets~\cite{Aaltonen:2011pg}}
\label{tauC7}
\end{figure}

Fig.~\ref{tauC7} shows that the angularity distribution indeed lies between the two expected limits. A good agreement is observed between the data and the PYTHIA sample. Furthermore, as in the jet mass case, an agreement is observed between the Midpoint results and those of anti-$k_T$.

\subsection{Planar Flow}
Planar flow~\cite{Almeida:2008tp, Almeida:2008yp} describes the way energy is deposited on the plane perpendicular to the jet axis. Due to the two-prong-nature of massive boosted QCD jets, their energetic signature in this plane is expected to be composed of two energy deposits, lying on a single line. This ideal configuration has a planar flow value of zero. The always existing soft contribution inside the jet shifts the most probable planer flow to a value somewhat higher than zero.\\
On the other hand, consider a hadronic decay of a boosted top quark, in which the $b$ quark and the $q\bar{q}$ pair are clustered together in a single jet. The three-pronged energy signature, with varying distances and energy distribution will yield a rather uniform planar flow distribution. This makes planar flow a good handle for separating two-prong decays from three-prong decays.

\begin{figure}[h]
\centering
\includegraphics[width=8cm]{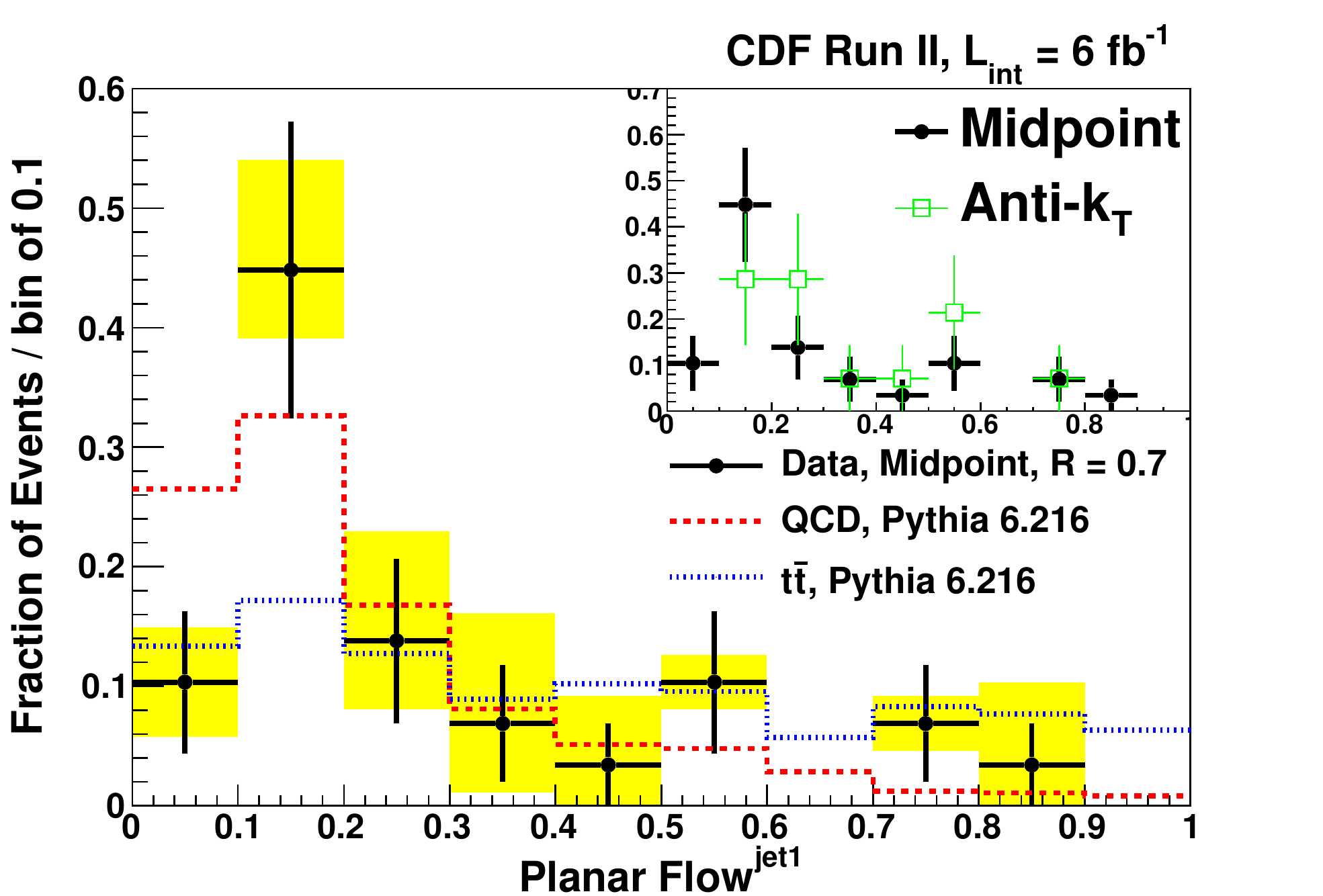}
\caption{The planar flow distributions for Midpoint jets with $p_T > 400~\rm{Gev}$,  $|\eta| \in (0.1, 0.7)$, and $m^{jet1} \in (130, 210)~\rm{Gev}^2$. Also shown are the pythia QCD (red dashed line) and  $t\bar{t}$ (blue dotted line) jets, as well as the results from the two jet algorithms (inset). All distributions have been separately normalized to unity~\cite{Aaltonen:2011pg}}
\label{pfC7}
\end{figure}

Fig.~\ref{pfC7} shows the planar flow distributions for jets in a mass window of 130 to 210 $\rm{Gev}^2$, a range relevant for searches of boosted tops. The data exhibits the expected QCD like behavior and peaks at a low planar flow value. This should be compared to the $t\bar{t}$ MC sample that exhibits a flatter distribution. Once again the two jet algorithms show good agreement. This is particularly interesting in the case of planar flow which deals with three-body configurations and considering the IR$_{3+1}$ ranking of Midpoint.

\subsection{Search for Boosted Tops}
Given the understanding of the jet substructure from the reviewed study, and in particular the jet mass, a technique was developed to separate the signal of boosted tops from the primary background to this search, which come from the production of massive QCD jets. Furthermore, a data-driven technique to estimate this background was used.

\begin{figure}[h]
\centering
\subfloat[Data]{\includegraphics[width=8cm]{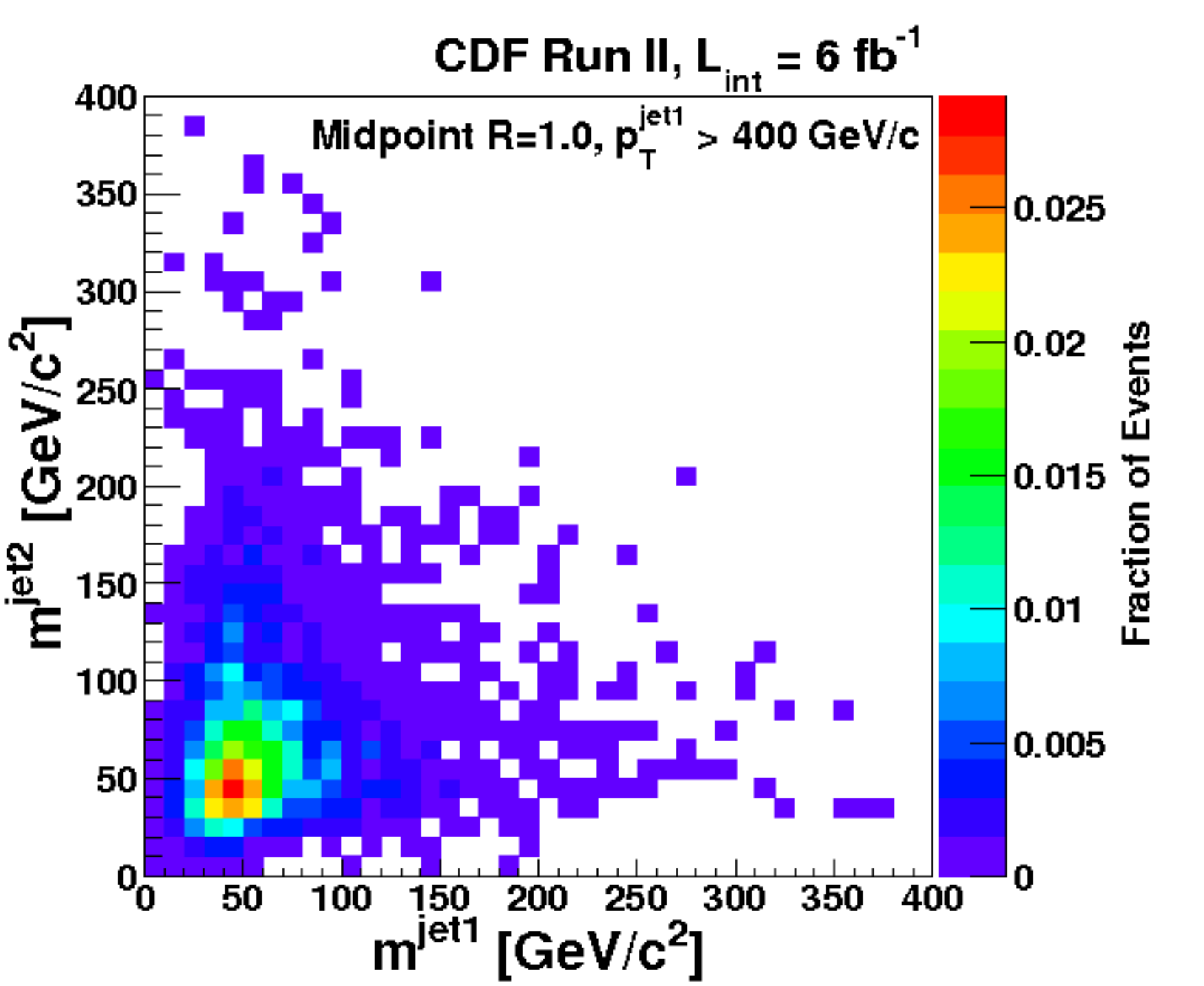}}
\subfloat[$t\bar{t}$ MC]{\includegraphics[width=8cm]{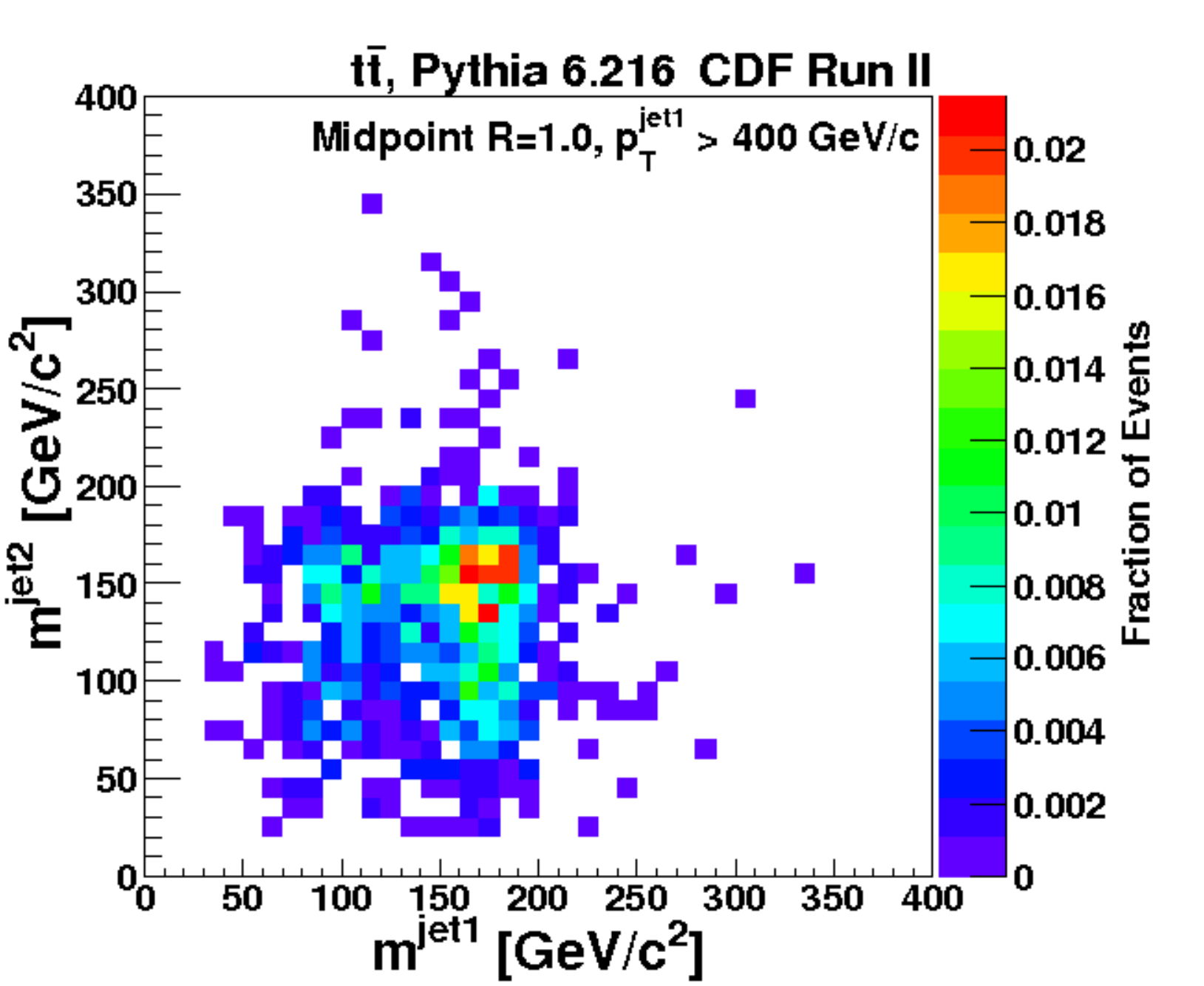}}
\caption{The $m^{jet2}$ versus $m^{jet1}$ distribution for data (a) and  $t\bar{t}$ MC (b) events with at least one jet with $p_T > 400~\rm{Gev}$ and $|\eta| < 0.7$ using $R = 1.0$ Midpoint cones. All events are required to have $S_{MET} < 4$~\cite{CDF:subjet2010b}}
\label{mjet2VSmjet1}
\end{figure}

In QCD, no dependence is expected in principle between the masses of the two leading jets in a typical dijet event. On the other hand, in all hadronic $t\bar{t}$ decays, one expects two massive jets recoiling one against each other. This is shown for the data and $t\bar{t}$ MC sample in Fig.~\ref{mjet2VSmjet1}. An upper cut on the significance of the missing energy was applied to reject semileptonic events.\\
The signal region, $D$, was defined to include two massive jets in the range of 130 to 210 $\rm{Gev}^2$. The definitions of the control regions and the number of counted events in each region is given in Table~\ref{Tab:MjetMjetEventCounts}. The table also shows the expected number of $t\bar{t}$ events in each region. The prediction shown in the table is given by:

\begin{eqnarray}
N_{D}^{pred} = \frac{N_B N_C}{N_A R_{mass}},
\label{dPrediction}
\end{eqnarray}
where $N_X$ is the number of events in region $X$ and $R_{mass} \equiv (N_B N_C)/(N_A N_D)$ was introduced in~\cite{Eshel:2011vs} and represents correlations that might exist between the $m^{jet2}$ and $m^{jet1}$ distributions. A value calculated using POWHEG~\cite{Alioli:2010xa, Nason:2004rx, Frixione:2007vw, Alioli:2010xd} of $R_{mass} = 0.89$ was used in this study.

\begin{table}[h]
    \center
      \leavevmode
      \begin{tabular}{|c|c|c|c|c|} \hline
    Region        &    $m^{jet1}$   &  $m^{jet2}$    &  Data  &  $t\bar{t}$\ MC       \\
            & ($\rm{Gev}^2$)  & ($\rm{Gev}^2$)  &  (Events)  & (Events) \\
         \hline  
   A    &    $(30,50)$  &  $(30,50)$  &   370  &  0.00 \\
   B    &    $(130,210)$  &  $(30,50)$  &  47  &  0.08 \\
   C   &    $(30,50)$    &   $(130,210)$ & 102  &  0.01  \\
   D (signal)  &  $(130,210)$  &  $(130,210)$  &  31  &  3.03  \\
   Predicted QCD in D  &   &   &  $14.6\pm 2.76$  &    \\
     \hline
   \end{tabular}
\caption{The observed number of events in the three control regions used to predict the background rate in the signal region (region $D$).  The $t\bar{t}$ MC event rates in each region are also shown~\cite{CDF:subjet2010b}}
\label{Tab:MjetMjetEventCounts} 
\end{table}  

\begin{figure}[h]
\centering
\subfloat[Data]{\includegraphics[width=8cm]{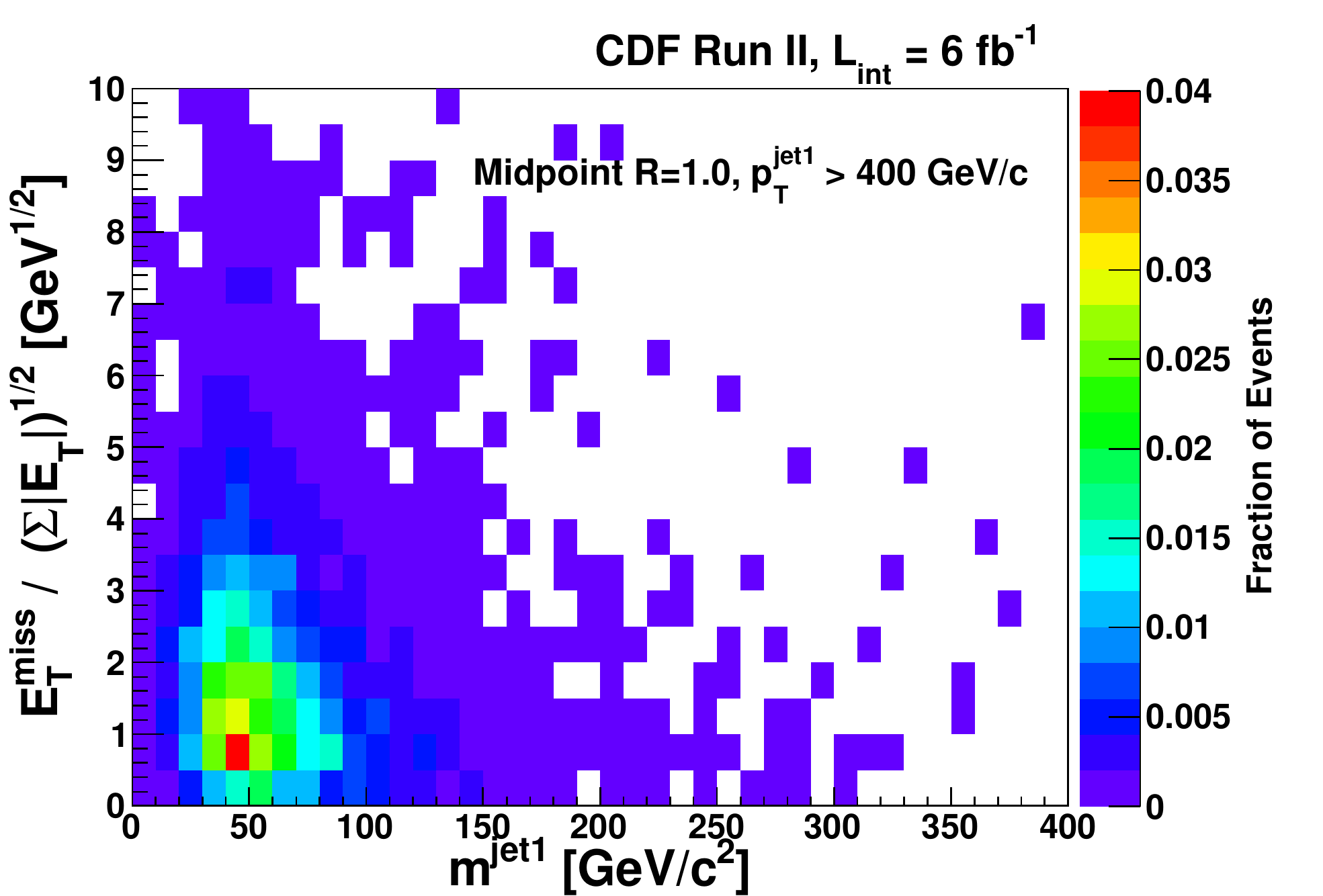}}
\subfloat[$t\bar{t}$ MC]{\includegraphics[width=8cm]{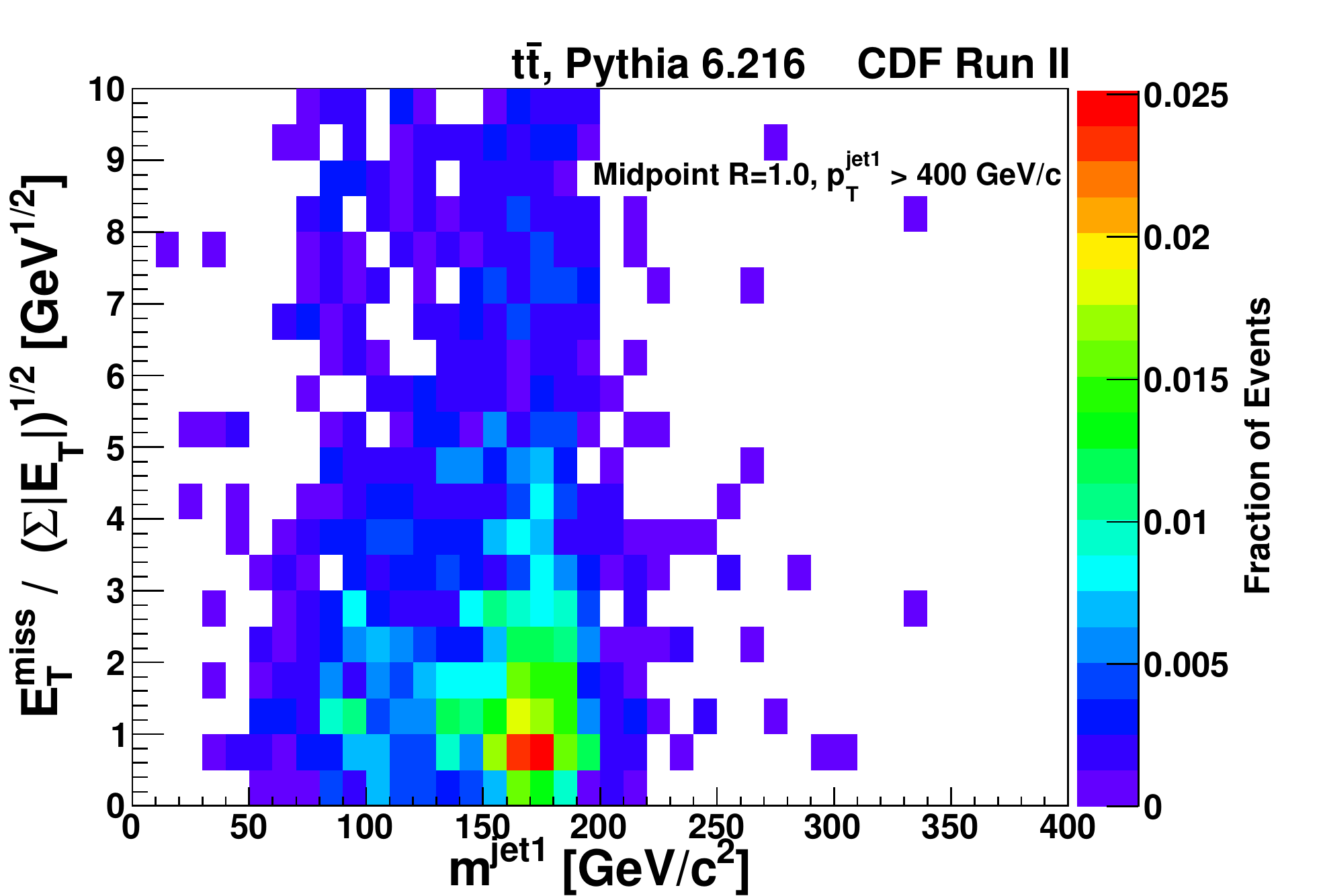}}
\caption{The $S_{MET}$ versus $m^{jet1}$ distribution for data (a) and $t\bar{t}$ MC (b) events with at least one jet with $p_T > 400~\rm{Gev}^2$ and $|\eta| < 0.7$ using $R = 1.0$ Midpoint cones~\cite{CDF:subjet2010b}}
\label{smetVSmjet1}
\end{figure}

In QCD events, no correlation is expected between the missing energy, which comes mostly from instrumental effects, and the leading jet mass. In semileptonic $t\bar{t}$ events, one expects a leading massive jet along with large missing energy. This is shown (using the missing energy significance) in Fig.~\ref{smetVSmjet1} for data and $t\bar{t}$ MC. The results are shown in Table~\ref{Tab:SigmetMjetEventCounts} as well as a prediction for the signal region obtained in a similar fashion to the prediction for the all hadronic case, omitting the $R_{mass}$ factor.

\begin{table}[htbp]
    \center
      \leavevmode
      \begin{tabular}{|c|c|c|c|c|} \hline
    Region        &    $m^{jet1}$   &  $S_{MET}$    &  Data  &  MC       \\
             & ($\rm{Gev}^2$)  & ($\sqrt{\rm{Gev}^2}$) &  (Events)  & (Events) \\
        \hline  
   A    &    $(30,50)$       &  $(2,3)$         & 256  &  0.01 \\
   B    &    $(130,210)$  &  $(2,3)$         &  42  &    1.07  \\
   C   &     $(30,50)$     & $(4,10)$      & 191  &  0.03   \\
   D (signal)  &  $(130,210)$  &  $(4,10)$  &  26  &  1.90   \\
   Predicted QCD in D  &   &   &  $31.3\pm 8.1$  &    \\
     \hline
   \end{tabular}
\caption{The observed number of events in the three control regions used to predict the background rate in the signal region (Region $D$) for high $S_{MET}$ and high $m^{jet1}$ .  The $t\bar{t}$ MC event rates are also shown normalized to the sensitivity of the data sample~\cite{CDF:subjet2010b}}
\label{Tab:SigmetMjetEventCounts} 
\end{table}  

A-priori a comparable signal-to-noise and acceptance in the all-hadronic and lepton+jets channels were expected. Thus the two channels were combined and used to set an upper limit on Standard Model $t\bar{t}$ production for top quarks with $p_T > 400~\rm{Gev}$. A 95\% C.L. limit was calculated, folding in the systematic uncertainties, using a Bayesian approach employing a flat prior on the cross section and treating the sources of systematic uncertainty as nuisance parameters~\cite{Amsler:2008zzb, PhysRevLett.103.092002}. The resulting upper limit is 38 fb at 95\% C.L.. This is approximately an order of magnitude higher than the estimated Standard Model rate, and is limited by the QCD background rates. It is, however, the most stringent limit on boosted top quark production to date.
Furthermore, the ``expected limit\rq{}\rq{} was calculated by using the background estimated from the data-driven technique and assuming an observation of $t\bar{t}$ events at the expected level of 4.9 events. The upper limit is 33 fb at 95\% C.L., which is lower than the observed limit since an excess of events above the expected signal plus background is observed in the data.
It is interesting to set a limit on the fully hadronic channel, as this creates a selection that is sensitive to pair production of two massive objects near the mass of the top quark. As the main interest in this case is in beyond-SM contributions to this final state, the background estimate the for the expected $t\bar{t}$ contribution of $3\pm0.8$ events is included. Taking out the top quark hadronic branching fraction of 4/9, the upper limit is 20 fb at 95\% C.L.

\section{Conclusions}
\label{sec:conclusion}

If beyond standard model physics is present  it should reveal itself in jet rich events. 
Understanding the substructure of such jets provides us with a handle in the analysis 
and differentiation of the possible models describing the events at current and future colliders.
Any machine probing the multi-TeV scale will produce electroweak scale resonances which will be highly boosted.
The methods we describe above should guide us and contribute in disentangling these resonances from overwhelming QCD
background.  Currently, these methods are being implemented on data to test their validity, and one such
 example is the CDF measurement.  At CDF the jet mass, angularity, and planar flow were measured for the
 first time for jets with $p_T > 400~\rm{Gev}$, using both Midpoint and anti-$k_T$ jet algorithms. A data-driven pileup correction 
 technique was developed and used. A good agreement between the data and PYTHIA MC predictions was observed. The jet mass 
 distribution for the data was consistent with the NLO QCD jet function at large mass. Angularity and planar flow show that the data 
 follows the two-prong assumption for boosted massive QCD jets. And interesting agreement between the two studied jet algorithms 
 is observed. Finally, a boosted top quark search based mainly on jet mas yielded an upper limit on the cross section of Standard 
 Model $t\bar{t}$ production at high $p_T$.

\acknowledgments
MS would like to thank all participants of the BOOST conferences for interesting discussions. 

\bibliographystyle{h-physrev}

\bibliography{subjet_review_bib}

\end{document}